\documentstyle[12pt]{article}
\def\lsim{\raisebox{-4pt}{$\,\stackrel{\textstyle{<}}{\sim}\,$}}
\def\gsim{\raisebox{-4pt}{$\,\stackrel{\textstyle{>}}{\sim}\,$}}
\begin{document}
\begin{flushright}
\baselineskip=12pt
CTP-TAMU-17/97\\
DOE/ER/40717--42\\
ACT-06/97\\
\tt hep-ph/9704247
\end{flushright}

\begin{center}
\vglue 1.5cm
{\Large\bf Compactifications of M-theory and their Phenomenological
Consequences\\}
\vglue 2.0cm
{\Large Tianjun Li$^{1,2}$, Jorge L. Lopez$^3$, and D.V. Nanopoulos$^{1,2,4}$}
\vglue 1cm
\begin{flushleft}
$^1$Center for Theoretical Physics, Department of Physics, Texas A\&M
University\\ College Station, TX 77843--4242, USA\\
$^2$Astroparticle Physics Group, Houston Advanced Research Center (HARC)\\
The Mitchell Campus, The Woodlands, TX 77381, USA\\
$^3$ Bonner Nuclear Lab, Department of Physics, Rice University\\ 6100 Main
Street, Houston, TX 77005, USA\\
$^4$ Academy of Athens, Chair of Theoretical Physics, Division of Natural
Sciences\\ 28 Panepistimiou Avenue, 10679 Athens, Greece\\
\end{flushleft}
\end{center}

\vglue 1.5cm
\begin{abstract}
We compactify the M-theory proposed by Horava and Witten on a Calabi-Yau
manifold with boundary $S_1/Z_2$. A no-scale-like K\"ahler potential, the
superpotential, and the gauge kinetic function are obtained in this
4-dimensional $E_6\times E_8$ model. We also study the general phenomenological
consequences of the resulting M-theory-inspired model, which may include very
light gravitinos, axions, and axinos.

\end{abstract}

\vspace{0.5cm}
\begin{flushleft}
\baselineskip=12pt
April 1997\\
\end{flushleft}
\newpage
\setcounter{page}{1}
\pagestyle{plain}
\baselineskip=14pt

\section{Introduction}
Recently, Horava and Witten~\cite{HW} presented a systematic analysis of
eleven-dimensional supergravity on a manifold with boundary, that is related to
the strong coupling limit of the $E_8\times E_8$ heterotic string. In this
novel theory, many previous successes based on conventional weakly-coupled
string theory may be preserved. In addition, Witten~\cite{Witten} offered an
explanation of why Newton's constant appears to be so small when the
4-dimensional grand unified gauge coupling ($\alpha_{\rm GUT}$) takes
experimentally acceptable values. It follows that the strengths of all
interactions, including gravitation, may be naturally unified at the GUT scale,
unlike the case of the weakly-coupled heterotic string. Many other interesting
implications have been studied, such as gluino condensation and supersymmetry
breaking~\cite{Horava}, the strong CP problem~\cite{BD}, threshold scale and
strong coupling effects~\cite{AQ,VK}, and phenomenological
consequences~\cite{LLN} which include a constrained sparticle spectrum within
the reach of present-generation particle accelerators.

In this paper we compactify this theory on a Calabi-Yau manifold with
Hodge numbers $h_{(1,1)}=1$ and $h_{(2,1)}=0$ and boundary $S_1/Z_2$. A
no-scale-like K\"ahler potential~\cite{no-scale}, the superpotential, and the
gauge kinetic function are obtained explicitly. In four dimensions this result
is related to the previous weakly-coupled string no-scale supergravity
result by Witten~\cite{Witten85} through a field transformation (Sec. 2),
which means that they are equivalent in four dimensions. One might then argue
that many heterotic string models obtained previously may also exist in the
M-theory regime.

In addition, we consider the physical couplings and scales in the Einstein
frame, the eleven-dimensional metric and fivebrane units, and give the
intermediate supersymmetry-breaking scale determined by the size of the
eleventh dimension of this theory~\cite{Witten,BD,AQ,VK,LLN,DM} for different
grand unified theories (Sec. 3). We also argue that there may exist a very
light gravitino in this scenario, which may explain the
$ee\gamma\gamma+E_{\rm T,miss}$ event~\cite{Park} observed by the CDF
Collaboration, and may have some further implications at LEP~2. Finally, we
comment on very light axions and axinos that might exist in this scenario.

\section{Formal derivation}
We follow the notation of Ref.~\cite{HW}, in which the bosonic part of the
eleven-dimensional supergravity Lagrangian is given by
\begin{eqnarray}
L_B&=&{1\over \kappa^2}\int_{M^{11}}d^{11}x\sqrt g
\left(-{1\over 2}R -{1\over 48}G_{IJKL}G^{IJKL}
\right.\left. -{\sqrt 2\over 3456}
\epsilon^{I_1I_2\dots I_{11}}C_{I_1I_2I_3}G_{I_4\dots I_7}G_{I_8\dots
I_{11}}\right)\nonumber\\
&&\qquad\qquad-\sum_{i=1,2}
{1\over\displaystyle 2\pi (4\pi \kappa^2)^{2\over 3}}
\int_{M^{10}_i}d^{10}x\sqrt g {1\over 4}F_{AB}^aF^{aAB}\,
\label{eq:1}
\end{eqnarray}
where $ G_{11\,ABC}=\left(\partial_{11}C_{ABC}\pm 23 \,\,{\rm
permutations}\right)
+{\kappa^2\over \sqrt 2 \lambda^2}\delta(x^{11})\omega_{ABC}$,
$\lambda^2=2\pi (4 \pi \kappa^2)^{2/3}$, and the gauge group at the boundary
is $E_8\times E_8$.

To perform the dimensional reduction to five dimensions (with the 4-dimensional
boundary where the Yang-Mills fields live) under the Calabi-Yau manifold with
Hodge numbers $h_{(1,1)}=1$ and $h_{(2,1)}=0$,
we follow Refs.~\cite{Witten85,FS} and keep only the SU(3) singlets in the
internal indices:
\begin{eqnarray}
g_{\mu \nu} \rightarrow g_{\mu \nu} ~;~ g_{i \bar j} = e^{\sigma}
\delta_{i \bar j} \,
\end{eqnarray}
\begin{eqnarray}
C_{\mu i \bar j}=iC_{\mu} \delta_{i \bar j} ~;~
C_{i j k}=C'\epsilon_{i j k} \,
\end{eqnarray}
Then, performing the Weyl rescaling:
\begin{eqnarray}
g_{\mu \nu} \rightarrow e^{-2\sigma} g_{\mu \nu} \,
\end{eqnarray}
and just paying
attention to the observable sector, which we assume is at the boundary
$x^{11}=0$,
we obtain the canonically normalized Einstein action:
\begin{eqnarray}
L_B&=&{V\over \kappa^2}\int_{M^5}d^5x\sqrt g
\left(-{1\over 2}R -{9\over 4}(\partial_{\mu}\sigma)^2-
{1\over 48}e^{6\sigma}G_{\mu \nu \rho \sigma} G^{\mu \nu \rho \sigma}
-27f_{\mu \nu}f^{\mu \nu}-36 e^{-3\sigma}|\partial_{\mu} {\hat C}'|^2
\right.\nonumber\\&&\left.
-54\sqrt 2 \epsilon^{\mu \nu \rho \sigma \delta} C_{\mu} f_{\nu \rho}
f_{\sigma \delta} + {3\over 4} \sqrt 2 i
\epsilon^{\mu \nu \rho \sigma \delta} \bar {{\hat C}'}
\stackrel{\leftrightarrow}{\partial_{\mu}} {\hat C}'
G_{\nu \rho \sigma \delta} \right)
\nonumber\\&&
-{V\over {2 \pi (4 \pi \kappa^2)^{2\over 3}}}\int_{M^4}d^4x\sqrt g
\left(-{1\over 4} e^{3\sigma}f Tr[F_{\mu \nu}F^{\mu \nu}]
-3 D_{\mu}C_x^* D^{\mu}C^x-{8\over 3} e^{-3\sigma}
|{{\partial W'}\over{\partial C}}|^2
\right.\nonumber\\&&\left.
-{9\over 2f} e^{-3\sigma}\sum_i (C^*, \lambda^i C)^2\right)\,
\end{eqnarray}
where
\begin{eqnarray}
W'&=& d_{xyz} C^x C^y C^z \,
\end{eqnarray}
\begin{eqnarray}
f_{5\mu} &=& \partial_5 C_{\mu} -\partial_{\mu} C_5 +
{i\over 6\sqrt 2}{\kappa^2\over \lambda^2} \delta (x^5) (C_x^*
\stackrel{\leftrightarrow}{D_{\mu}}C^x) \,
\end{eqnarray}
\begin{eqnarray}
\partial_{\mu} {\hat C}' &=& \partial_{\mu} C' +
{\sqrt{2}\over 3} \delta_{\mu 5}
\delta(x^5)
{\kappa^2\over \lambda^2} W' \,
\end{eqnarray}
The definitions of the $C_x$, $d_{xyz}$ and $f$ are the same as in the
Ref.~\cite{Witten85}, and $V$ is the coordinate volume of the Calabi-Yau
manifold, i.e., $V=\int d^6x$. In addition, the (observable) gauge group at the
boundary ($x^{11} =0$ or $x^5 =0$) is now $E_6$ with spin connection
embedding, as in Ref.~\cite{Witten85}.

If we define a pseudoscalar $D$ by a duality transformation:
\begin{eqnarray}
{1\over 4!}e^{6\sigma} G_{\mu \nu \rho \sigma} &=&
\epsilon_{\mu \nu \rho \sigma \delta} ({\partial^{\delta}D+{3\over 4}\sqrt 2
i \bar {{\hat C}'}
\stackrel{\leftrightarrow}{\partial^{\delta}} {\hat C}'}) \,
\end{eqnarray}
we have the following K\"ahler potential in the five-dimensional bulk:
\begin{eqnarray}
K &=& -\ln\,[S+\bar S-72\bar {C'} C'] \,
\end{eqnarray}
where
\begin{eqnarray}
S &=& e^{3\sigma}+i24\sqrt 2 D +36\bar {C'} C' \,
\end{eqnarray}
This K\"ahler potential parametrizes the ${SU(2,1)\over {SU(2)\times SU(1)}}$
quaternionic manifold~\cite{FS,CCAF}.
Also, $D$ is the invisible axion.

We now compactify the above 5-dimensional-with-boundary Lagrangian on
$S_1/Z_2$. For the fields with 11-dimensional origin, i.e.,
the fields in the bulk, we keep only the zero modes, and considering the
boundary condition we can expand $\delta (x^5)$ as:
\begin{eqnarray}
\delta(x^5)&=&{1\over {2\pi \rho}} + {1\over {\pi \rho}}
\sum_{n=1}^{\infty} \cos{{nx^5}\over \rho}\,
\end{eqnarray}
where $\rho$ is the coordinate radius of $S_1$, i.e.,
$\rho ={1\over {2\pi} }\int dx^5$.
Furthermore, choosing the following elfbein form:
\begin{eqnarray}
e_M^A &=& \left|\matrix{e_{\mu}^a & 0 \cr
0 & \varphi \cr}\right| \,
\end{eqnarray}
and performing the Weyl rescaling:
\begin{eqnarray}
g_{\mu \nu} \rightarrow \varphi^{-1} g_{\mu \nu} \,
\end{eqnarray}
we obtain the 4-dimensional Lagrangian in the Einstein frame (note that at the
boundary $C'$=0 and $C_{\mu \nu \rho} = 0$):
\begin{eqnarray}
L_B&=&{V\over \kappa^2} {2 \pi \rho} \int d^4x\sqrt g \left(
-{1\over 2}R-{1\over 12} e^{6\sigma} G_{11\mu \nu \rho} G^{11\mu \nu \rho}
-{9\over 4} (\partial_{\mu} \sigma)^2
-{3\over 4} ({{\partial_{\mu} \varphi}\over \varphi})^2
\right.\nonumber\\&&\left.
-54 \varphi^{-2}(\partial_{\mu} C_5-{i\over {6\sqrt 2}}{1\over {2\pi \rho}}
{\kappa^2 \over \lambda^2} C_x^*
\stackrel{\leftrightarrow}{D_{\mu}}C^x)^2
\right.\nonumber\\&&\left.
-{1\over {2\pi \rho}} {\kappa^2 \over \lambda^2}
[ 3\varphi^{-1}D_{\mu}C_x^*D^{\mu}C^x+
{1\over 4}fe^{3\sigma}{\rm Tr}\,[F_{\mu \nu}F^{\mu \nu}]
+{8\over 3} \varphi^{-2}e^{-3\sigma}
|{{\partial W'}\over {\partial C}}|^2
\right.\nonumber\\&&\left.
+{9\over 2f} \varphi^{-2}e^{-3\sigma}\sum_i
(C^*, \lambda^i C)^2]
-8({1\over {2\pi \rho}})^2 ({\kappa \over \lambda})^4e^{-3\sigma}
\varphi^{-3} |W'|^2\right)
\label{eq:15}\\
\end{eqnarray}
Finally, if we define $g_c^2=2 \pi \rho \lambda^2/\kappa^2$ and perform the
transformation:
\begin{eqnarray}
A_{\mu} \rightarrow g_c A_{\mu} ~;~C_x \rightarrow g_c C_x \,
\end{eqnarray}
we obtain the standard supergravity Lagrangian:
\begin{eqnarray}
L_B&=&{V\over \kappa^2} {2 \pi \rho} \int d^4x\sqrt g \left(
-{1\over 2}R-{1\over 12} e^{6\sigma} G_{11\mu \nu \rho} G^{11\mu \nu \rho}
-{9\over 4} (\partial_{\mu} \sigma)^2
-{3\over 4}({{\partial_{\mu} \varphi}\over \varphi})^2
\right.\nonumber\\&&\left.
-54 \varphi^{-2}(\partial_{\mu} C_5-{i\over {6\sqrt 2}}
C_x^*
\stackrel{\leftrightarrow}{D_{\mu}}C^x)^2
\right.\nonumber\\&&\left.
-3\varphi^{-1}D_{\mu}C_x^*D^{\mu}C^x
-{1\over 4}fe^{3\sigma}Tr[F_{\mu \nu}F^{\mu \nu}]
-{8\over 3} g_c^2 \varphi^{-2}e^{-3\sigma}
|{{\partial W'}\over {\partial C}}|^2
\right.\nonumber\\&&\left.
-{9\over 2f}g_c^2 \varphi^{-2}e^{-3\sigma}\sum_i
(C^*, \lambda^i C)^2
-8g_c^2e^{-3\sigma}\varphi^{-3} |W'|^2\right)\nonumber\\
\end{eqnarray}

{}From this expression, neglecting the overall factor ${V\over \kappa^2} 2\pi
\rho$, and defining the pseudoscalar by the duality transformation:
\begin{eqnarray}
{1\over 4!} e^{6\sigma } G_{11\mu \nu \rho} &=&
\epsilon_{\mu \nu \rho \sigma} (\partial^{\sigma} D) \,
\end{eqnarray}
we obtain the following K\"ahler potential:
\begin{eqnarray}
K &=& -\ln\,[S+\bar S]-3\ln\,[T+\bar T-2 C_x^* C^x]  ~,~ \,
\end{eqnarray}
where
\begin{eqnarray}
S=e^{3\sigma}+i24\sqrt 2 D \,
\end{eqnarray}
and
\begin{eqnarray}
T=\varphi -i6\sqrt 2 C_5
+ C_x^*C^x \,
\end{eqnarray}
Here $D$ and $C_5$ are the pseudoscalars and the invisible
axions~\cite{Witten85}. In addition, we have the following gauge kinetic
function:
\begin{eqnarray}
Ref_{\alpha \beta} &=& f ReS\, \delta_{\alpha \beta} \,
\end{eqnarray}
and the superpotential $W$,
\begin{eqnarray}
 W= 8\sqrt {2\over 3} g_c\, d_{x y z} C^x C^y C^z \,
\end{eqnarray}
Furthermore,
if we perform the field transformation:
\begin{eqnarray}
\varphi \rightarrow \phi^{3/4} e^{\sigma} ~;~e^{\sigma} \rightarrow
\phi^{-1/4} e^{\sigma} \,
\end{eqnarray}
we obtain $S, T$ fields which are similar to the previous result in
Ref.~\cite{Witten85}, as noticed in the Ref.~\cite{CCAF}. This is an
interesting result, and allows us to argue that the weakly-coupled heterotic
string models derived previously may exist in the M-theory proposed by Horava
and Witten. The above 4-dimensional results should be related to the
4-dimensional results from the weakly-coupled heterotic string compactification
by a field transformation, although we note that there may not exist a
10-dimensional effective field theory (EFT)~\cite{VK}.

\section{Phenomenological consequences}
Let us now discuss the physical couplings and the physical radius of the
eleventh dimension ($\rho_p$) in the various frames. If we define the Planck
mass in $d$ dimensions as $8 \pi G_{N}^{(d)}= M_d^{2-d}=\kappa^2_d$ and
$M_4=M_{\rm Pl}=2.4\times 10^{18}$ GeV,
from the 11-dimensional Lagrangian [Eq.~(\ref{eq:1})],
we have $M_{11}=\kappa^{-2/9}$, and from  the above 4-dimensional Lagrangian
[Eq.~(\ref{eq:15})] in the Einstein frame, we obtain:
\begin{eqnarray}
8\pi\, \left[G_{N}^{(4)}\right]_E &=& {\kappa^2 \over {2\pi \rho V}}\,\\
\left[\alpha_{\rm GUT}\right]_E &=& {1\over{2 V_p f}}\,(4\pi \kappa^2 )^{2/3}\,
\end{eqnarray}
where $V_p$ is the physical volume of the Calabi-Yau manifold, i.e., $V_p = V
ReS $. Similar relations have been obtained by Witten and
others~\cite{Witten,BD,AQ,VK} in the metric of the eleven-dimensional theory,
these are:
\begin{eqnarray}
8\pi\,\left[G_{N}^{(4)}\right]_W &=& {\kappa^2 \over {2\pi \rho_p V_p}} \ , \\
\left[\alpha_{\rm GUT}\right]_W &=&{1\over {2 V_p f}}\,(4\pi\kappa^2 )^{2/3}\,
\end{eqnarray}
where $\rho_p$ is physical radius, i.e., $\rho_p={1\over {2\pi}} \int
dx^{11} {\sqrt {g_{11, 11}}}$. We have also included the constant $f$,
which arises from the following normalization: if we asssume that the
4-dimensional grand unified group is $G$ (i.e., $E_8$ is broken to G) and if
$T$ is a generator of $G$, ${\rm Tr}_{E_8}$ and ${\rm Tr}_{G}$ are traces in
the adjoint representations of $G$ and $E_8$, then ${\rm Tr}_{E_8} T^2 = f\,
{\rm Tr}_G T^2$ \cite{Witten85}.

Because eleven-dimensional supergravity can be derived from the
eleven-dimensional supermembrane world volume action by imposing kappa symmetry
\cite{Sezgin}, and there are arguments in favor of a membrane/fivebrane duality
in eleven dimensions \cite{Duff}, it has been argued that the fivebrane units
are the natural or fundamental units of M-theory \cite{DM}. Therefore, we also
consider the above relations in the fivebrane units~\cite{DM, PBIN}. We
continue to use our above notation, i.e., we do not use the membrane
quantization condition~\cite{DM},
since we want to obtain an explicit expression for
$\rho^{-1}_p$ and the eleven-dimensional fundamental constant $\kappa$;
the result is the same in both approaches. We obtain the
relevant 4-dimensional Lagrangian in fivebrane units~\cite{DM,PBIN}:
\begin{eqnarray}
L_5 &=&-\int d^4 x ({1\over {2\kappa^2 }} 2 \pi \rho V e^{\sigma}
e^{2\phi /3} R + {f\over {2\pi }} (4\pi \kappa^2)^{-2/3} V e^{3\sigma}
{\rm tr}\,F_{\mu \nu} F^{\mu \nu})\,
\end{eqnarray}
therefore we have:
\begin{eqnarray}
8\pi\left[G_{N}^{(4)}\right]_{5B} &=&
{\kappa^2 \over {2\pi \rho_p V_p}} e^{2\sigma } \ , \\
\left[\alpha_{\rm GUT}\right]_{5B} &=& {1\over{2V_pf}}\,(4\pi \kappa^2)^{2/3}\,
\end{eqnarray}
where $V_p = e^{3\sigma} V$ and $\rho_p =\rho e^{2\phi /3}$. We note that in
all three cases (Einstein, Witten, and fivebrane), $\alpha_{\rm GUT}$ is the
same, as the term $\sqrt g\, {\rm tr}\,F^2$ is invariant under the rescaling of
$g_{\mu \nu}$ in four dimensions.
Therefore, we have in general:
\begin{eqnarray}
M_{11} &=& \left[2 (4\pi )^{-2/3}\, V_p f\, \alpha_{\rm GUT}\right]^{-1/6}\,
\end{eqnarray}
If we define $V_p = L^d l^{6-d}$, ( $0 \leq d \leq 6$ ), where $L^{-1}$ is the
compactification scale and $l$ is the small internal length, we obtain:
\begin{eqnarray}
L^{-1} &=& \left[2(4\pi )^{-2/3} f \alpha_{\rm GUT}\right]^{1/d}
\left({M_{11}\over l^{-1}}\right)^{(6-d)/d}\, M_{11} ~,~ \,
\end{eqnarray}
which tells us that $L^{-1}\sim M_{11}$, when $l^{-1} \sim M_{11}$.

Let us now discuss $\rho^{-1}_p$ in the eleven-dimensional metric and fivebrane
units (it is not natural to think of the Lagrangian in the Einstein frame as
fundamental). We obtain:
\begin{eqnarray}
\left[\rho^{-1}_p\right]_W &=& 8 \pi^2 \left(2 f\alpha_{\rm GUT}\right)^{-3/2}
\left(M_{\rm Pl}^W\right)^{-2} V_p^{-1/2}\, \\
\left[\rho^{-1}_p\right]_{5B} &=& 8\pi^2\left(2 f\alpha_{\rm GUT}\right)^{-3/2}
\left(M_{\rm Pl}^{5B}\right)^{-2}  V_p^{-1/2} e^{-2\sigma }
\end{eqnarray}
The eleven-dimensional length $(\pi \rho_p)^{-1}$, is of great phenomenological
importance because it is related to the scale of supersymmetry breaking
\cite{Horava, LLN}. To obtain numerical results we set $M_{Pl}^W = M_{Pl}^{5B}
= M_{Pl}= 2.4\times 10^{18}\,{\rm GeV}$, $\alpha_{\rm GUT}= {1\over 25}$,
$V_p=M_{\rm GUT}^{-6}$, and $M_{\rm GUT} = 10^{16}\,{\rm GeV}$. We also set
$f=1$ for simplicity and to facilitate comparison with previous papers which
only consider this case. We find
$[(\pi \rho_p)^{-1}]_W\sim 1.9\times 10^{14}\,{\rm GeV}$ and
$[(\pi \rho_p)^{-1}]_{5B}\sim 1.2\times 10^{13}\,{\rm GeV}$ (as
$e^{2\sigma}\approx16.6$ if we set
$V^{-1/6}=M_{11}$).

We now relax the $f=1$ choice and consider the case of realistic grand
unified groups. The results for $[(\pi \rho_p)^{-1}]_W$ and $[(\pi
\rho_p)^{-1}]_{5B}$ for $G=E_6$, SO(10), SU(5), and SU(5)$\times$U(1)
are listed in Table~\ref{Table1}. (We have assumed $V^{-1/6} = M_{11}$ in all
these cases.) We then generally conclude that the supersymmetry-breaking scale
is expected to be in $(10^{12}-10^{14})\,{\rm GeV}$ range.

\begin{table}[t]
\caption{Supersymmetry-breaking scales in the eleven-dimensional metric (W) and
fivebrane units (5B) for various choices of (observable) unified gauge groups
($G$). The parameter $f$ defined in the text is also listed in each case. All
scales in GeV.}
\label{Table1}
\begin{center}
\begin{tabular}{c|c|c|c}
$G$&$f$&$[(\pi\rho_p)^{-1}]_W$&$[(\pi\rho_p)^{-1}]_{5B}$\\ \hline
$E_6$&2.5&$4.9\times10^{13}$&$5.3\times10^{12}$\\
SO(10)&3.75&$2.6\times10^{13}$&$3.8\times10^{12}$\\
SU(5)&6&$1.3\times10^{13}$&$2.6\times10^{12}$\\
SU(5)$\times$U(1)&6&$1.3\times10^{13}$&$2.6\times10^{12}$\\
\end{tabular}
\end{center}
\hrule
\end{table}

Before addressing further phenomenological features of this scenario, we would
like to connect up with our previous phenomenologically oriented study of
M-theory--inspired no-scale supergravity in Ref.~\cite{LLN}. In that paper we
assumed a supergravity model with a no-scale-supergravity--like K\"ahler
potential (implying vanishing universal scalar masses $m_0=0$), as suggested by
earlier work in Refs.~\cite{Horava,BD}. In the present paper we have shown
explicitly that such assumption is justified. Moreover, the transmission of
supersymmetry-breaking effects from the hidden to the observable sector was
assumed to follow the mechanism outlined by Horava \cite{Horava}, whereby such
effects are only felt for scales below $(\pi \rho_p)^{-1}$. Our previous
sampling of such scales agrees well with our present results in
Table~\ref{Table1}. These two ingredients were shown \cite{LLN} to lead to a
rather restricted spectrum of superparticle masses within the reach of the
present generation of accelerator experiments. It is also worth pointing out
another, perhaps more intuitive, explanation of the $m_0=0$ result: since in
M-theory the observable sector fields live in the twisted sector of an
orbifold, general arguments \cite{AMQ} indicate that such fields should not
feel any supersymmetry-breaking effects.

Because the supersymmetry breaking scale is low, there may exist light
gravitinos in the M-theory regime. As has been explored in detailed recently,
such light gravitinos may explain the $ee\gamma\gamma+E_{\rm T,miss}$ event
reported by the CDF Collaboration, and might have some interesting consequences
at LEP2 \cite{JLDN}. In this connection, it was noticed some time ago that
reactions such as  $gg\to \tilde g\widetilde G,\widetilde G\widetilde G$
(where $g,\tilde g,\widetilde G$ stand for gauge boson, gaugino, and gravitino
respectively) may exceed the tree-level unitarity limit because of
the non-renormalizability of the low-energy effective gravity theory
\cite{Roy}. The critical energy was estimated to be $E_{cr} \sim c M_{\rm Pl}\,
m_{3/2}/m_{\tilde g}$, with $c\sim10^2$. As tree-level unitarity is violated
for $E>E_{cr}$, above $E_{cr}$ the theory is expected to become strongly
interacting or change its structure (or both) in order to restore unitarity.
The eleventh dimension threshold $(\pi\rho_p)^{-1}$ appears to fulfill such
requirements, and thus one might require $(\pi\rho_p)^{-1}<E_{cr}$, implying
a lower bound on $m_{3/2}$ for any given supersymmetry-breaking scale. Taken at
face value, this constraint and the results in Table~\ref{Table1} appear to
require $m_{3/2}>(10^2-10^4)\,{\rm eV}$, which may be consistent with the
estimated $m_{3/2}<250\,{\rm eV}$ required for the $\chi\to\gamma+\widetilde G$
decay to occur within the detector \cite{JLDN}. However, as we are not sure
whether the above unitarity constraints can be used directly in M-theory (as
they depend on specific processes that might be forbidden in this scenario), we
find the above level of consistency rather encouraging.

There also exist axions in this scenario, for example $D$: if we define
$e^{6\sigma} \partial_{\mu} C_{11\nu \rho } =\epsilon_{\mu \nu \rho \sigma }
\partial^{\sigma} \theta $, we have a ($\theta F^{\mu \nu } {\tilde{F}}_{\mu
\nu }$) term from ($G_{11\mu \nu \rho} G^{11\mu \nu \rho}$). The strong CP
problem may be solved by such axions~\cite{BD}. In this scenario,
because the decay constant of these axions is very high $\sim 10^{16}$ GeV
\cite{BD, KMCI}, their axino superpartners, if they are very light, might not
provide the alternative explanation to the CDF missing energy event proposed in
Ref.~\cite{HKY}.

\section{Conclusions}
We have constructed an explicit 4-dimensional $E_6\times E_8$ model from the
M-theory of Horava and Witten. We have also calculated the physical couplings
and physical eleventh dimension length, which is presumed to be related to the
supersymmetry breaking scale. We have done these calculations in various
metrics and shown how they are all related to each other. Finally we discussed
the phenomenological consequences of such scenario, which may include light
gravitinos, axions, and axinos.

\section*{Acknowledgments}
T. Li would like to thank K. Benakli, J. X. Lu, and C. N. Pope for useful
discussions. This work has been partially supported by the World Laboratory.
The work of J.~L. has been supported in part by DOE grant DE-FG05-93-ER-40717
and that of D.V.N. by DOE grant DE-FG05-91-ER-40633.

\end{document}